\documentclass[preprint,12pt,num,sort&compress]{elsarticle}

\usepackage{amssymb,amsmath}
\usepackage{xcolor,hyperref}
\usepackage{slashed}
\usepackage{wrapfig}

\journal{Physics Letters B}
\hypersetup{colorlinks=true, citecolor=magenta, urlcolor=magenta, linkcolor=magenta}
\begin{document}
\flushbottom

\begin{frontmatter}



  \title{The Fundaments of Unity: ${\mathcal O}(1)$ Dimensionless Couplings in Quantum Field Theories} 
  
  \author[1]{Ben Allanach}
  \ead{ben.allanach.work@gmail.com}
  \affiliation[1]{organization={DAMTP, University of Cambridge},
    addressline={Centre for Mathematical Sciences, Wilberforce Road},
    postcode={CB3 0WA},
    city={Cambridge},
    country={United Kingdom}}

  \begin{abstract}
We critically examine the expectation that in a fundamental quantum field theory, dimensionless couplings in the Lagrangian density should all be of order unity. We propose a measure to quantify the adherence of a theory to this: the spread (the ratio of the largest to the smallest of the magnitudes) of such dimensionless couplings, obtaining various closed-form results. If we take independent and identically distributed (IID) couplings to parameterise our uncertainty on the values of the order-unity couplings, the spread can be much larger than one might naively expect. For a theory with 20 IID unit normal couplings, the probability that the spread is greater than 100 is 0.29, for example. Even when the IID couplings have exponentially suppressed tails, the distribution of the spread has fat \emph{power-law} tails whose amplitude grows with the number of independent couplings. 
  \end{abstract}

  
\end{frontmatter}

A recurring expectation of a fundamental theory  is that its Lagrangian is controlled
by dimensionless\footnote{We work in natural units where
the speed of light $c$ and $\hbar$ are set to unity and are dimensionless.} couplings of order 
unity~\cite{TOE}. Here, we wish to address this pervasive but rarely quantified assumption in model building.
%
We consider a fundamental quantum field theory in $3+1$ dimensional spacetime at
a high energy scale: just below the Planck scale or the string scale, for
example. 
The dimensionless couplings seen at lower energies in an effective quantum
field theory 
would then be fixed in terms of their values at the high scale
through calculable order unity factors such as renormalisation-group
running and perhaps group-theoretic (Clebsch--Gordan) coefficients.
Clebsch--Gordan coefficients, being
the coefficients in the decomposition of 
tensor products of irreducible representations, are of order unity provided
the 
number of quantum fields --- which populate the representation space of said
representations --- 
is not huge. 

In the case where there is unification in the theory, it could be that the
number of \emph{independent} high-scale dimensionless couplings is lower than
their number in the effective field theory: some of the high-scale couplings may
be related (equal in the case of gauge unification), for example.
In the extreme case, there is only one independent high-scale dimensionless
coupling, whereby all
low-energy dimensionless couplings are calculable from it. In this paper, we consider
the other cases where there is more than one independent dimensionless
coupling. In many string theory models, for example, despite there being a
single string coupling strength, 
the values of dimensionless
couplings in the quantum field theory valid just below the string scale are
fixed by the vacuum expectation values of various moduli
fields~\cite{Witten:1985xb,Kaplunovsky:1995jw}. This high-scale quantum field
theory then effectively possesses independent dimensionless couplings (at
least until a vacuum and compactification of the string model in question is
solved). 

The dimensionless couplings measured in the Standard Model (SM) are, however, far from
being all of order unity. The Yukawa couplings of charged fermionic fields span
more than five orders of magnitude (the `fermion mass puzzle'). 
Such hierarchies sit uneasily with the order-unity expectation and are a standard
motivation for physics beyond the SM\@.
The fermion mass puzzle can be
solved by, for example, a more fundamental high-scale theory as exemplified by
Froggatt--Nielsen (FN)-type models~\cite{Froggatt:1978nt}.
The hierarchies in the fermion masses are then given by
various positive
powers of suppression factors (of order $0.1$) in effective non-renormalisable
operators. The order $0.1$ number is the ratio of the vacuum expectation value
of a field breaking a $U(1)$-FN gauge symmetry to the mass of a heavy fermion in
a vector-like representation of the Lie algebra of the gauge group.
In FN models, there is typically a greater number of
order unity couplings in the more fundamental
high-scale theory than
there are effective Yukawa couplings in the low-scale effective field theory. 

It is our purpose to quantify a measure of how far a general quantum field theory
is from having order-unity couplings and to examine some implications.
This measure could then be used to quantify and rank how successful various 
models are in achieving the order-unity expectation.  

We shall treat the
dimensionless couplings as IID random 
variables whose prior probability distribution encodes our uncertainty about
their values. Bayesian statistics will then provide a natural quantification
of the probability distribution of the amount of spread in their values. 

Over a set of IID dimensionless couplings $\{Y_i\}$ (where $i$ labels the
coupling in question) each distributed according to 
a probability density function (PDF) $p(x)$, we define the quantity
\begin{equation}
  R:=\max_i(\{|Y_i|\})/\min_i(\{|Y_i|\}) \geq 1. 
\end{equation}
$R$ then quantifies the amount of spread between the magnitudes of the dimensionless
couplings. It gives a statistic which, given an assumption for $p(x)$ and a
number of dimensionless couplings, has a PDF which is calculable. 
We shall initially ignore imaginary parts of the dimensionless couplings. The
resulting complex phases could be considered to be separate
parameters which multiply $|Y_i|$, in which case our arguments below
remain for the $|Y_i|$,
or we could double the number of $Y_i$, counting half of them as real parts
and half as imaginary parts. Either way, we do not expect any significant
change to the conclusions of the first part of our analysis.
The imaginary parts of the couplings are however important when
considering particle mixing and so we shall consider $3 \times 3$ matrices of
complex-valued dimensionless couplings. This is of particular relevance to
Yukawa couplings, which we know to be complex because of various measurements
of $CP$ violation~\cite{PDG}. 

Consider first a theory with just two real IID couplings $Y_1$ and $Y_2$, leaving
$p(x)$ unspecified for now. Their spread is
\begin{equation}
  R=\frac{\max(|Y_1|,|Y_2|)}{\min(|Y_1|,|Y_2|)}
   =\max\!\left(\left|\frac{Y_1}{Y_2}\right|,\left|\frac{Y_2}{Y_1}\right|\right) .
\end{equation}
It is convenient to first compute the distribution of the signed ratio
$t:=Y_1/Y_2$. Changing variables from $(Y_1,Y_2)$
to $(t,Y_2)$ in the normalised joint distribution $p(Y_1,Y_2)=p(Y_1)p(Y_2)$
yields a Jacobian factor $|Y_2|$,
\begin{equation}
  p(t)=\int_{-\infty}^{\infty} dY_2\ |Y_2|\,p(Y_2 t,\,Y_2). \label{p_of_R}
\end{equation}

\begin{figure}
  \begin{center}
    \includegraphics[width=0.7\textwidth]{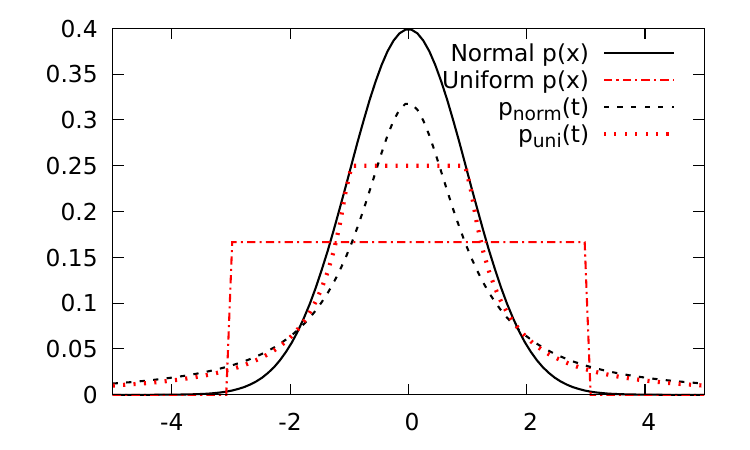}
  \end{center}
  \caption{\label{fig:dists} The distribution in $t=Y_1/Y_2$ for two
    order-unity IID
    couplings $Y_1$ and $Y_2$. The normal distribution is shown along with
    $p_{\text{norm}}(t)$, the associated distribution in $t$. The uniform distribution is shown
    along with $p_{\text{uni}}(t)$, the associated distribution in $t$. 
    Both ratio tails fall as $1/t^2$.}
\end{figure}
To make further progress, we should be more explicit about a form of $p(x)$. 
Let us take $p(x)$ to be a unit normal (also known as the unit Gaussian) centred on zero,
$p(x)=e^{-x^2/2}/\sqrt{2\pi}$, shown as the unbroken line in Fig.~\ref{fig:dists}.
The unit normal encodes the order-unity assumption: its magnitude is very
likely 
not greater than about 10 (the probability is $1.5\times 10^{-23}$), for example.
We shall consider a different (but
arguably as feasible) distribution later, to explore how much this choice
matters. 
The joint distribution of $Y_1$ and $Y_2$ is then, by the IID assumption,
just the product of $p(Y_1)$ and $p(Y_2)$:
\begin{equation}
  p(Y_1, Y_2) = \frac{1}{2\pi} \exp \left( - \frac{Y_1^2+Y_2^2}{2}\right) \label{joint}
  \end{equation}
Substituting (\ref{joint}) into (\ref{p_of_R})
gives
\begin{equation}
  p_{\text{norm}}(t)=\frac{1}{2\pi}\int_{-\infty}^{\infty}dY_2\,|Y_2|\,e^{-Y_2^2(1+t^2)/2}
  =\frac{1}{\pi(1+t^2)}, \label{p_N}
\end{equation}
the Cauchy distribution\footnote{The fact that the ratio of two zero-centred
normal distributions is Cauchy distributed was first shown in
Ref.~\cite{Geary:1930}.}, plotted in Fig.~\ref{fig:dists}. 
The Cauchy distribution has fatter tails than
does the normal distribution\footnote{The Cauchy
distribution has no well-defined mean or variance. This means, for example, that
estimates of the mean of 
$R$ using Monte Carlo evaluation will not converge. 
The Cauchy distribution does have a well-defined PDF so that intervals
can be properly defined, however.}. 
We have thus arrived at our central result: that although we started with
couplings that have 
exponentially suppressed tail distributions, in their ratio, we end up with
only a \emph{power suppressed} tail. 

One may ask whether such fat tails are due to us picking the ratio of
\emph{normal} distributions in the couplings; after all, the denominator of
the ratio is
peaked at zero. 
We might well instead have restricted the dimensionless couplings by the requirement
of perturbativity. 
Perturbativity (i.e.\ accuracy of a perturbative expansion in terms of quantum
field theoretic loops)
is lost when the magnitude of a dimensionless coupling becomes of order the
inverse of the square root of a loop 
factor, i.e.\ $4 \pi$. The exact value where one loses perturbativity is 
not precisely defined, but 
if one considers the
ratio of two IID
variates each uniformly distributed between $-3$ and $3$ but zero elsewhere (such that there is
no further enhanced peak around zero and there is no chance that 
the magnitude of either the denominator or the numerator is
larger than $4\pi$), an analogous computation to that in
(\ref{joint})--(\ref{p_N}) yields
\begin{equation}
  p_\text{uni}(t) = \left\{\begin{array}{cc}
  \frac{1}{4}    & |t| \leq 1 \\
  \frac{1}{4t^2} & |t|    > 1 \\ 
  \end{array}
  \right. . \label{p_U}
  \end{equation}
We plot $p_{\text{uni}}(t)$ in Fig.~\ref{fig:dists}, where the distribution can be
compared to $p_{\text{norm}}(t)$.
We see from (\ref{p_U}) and the figure that at large $|t|$, the distribution
drops proportional to $1/t^2$, as does the ratio distribution coming from unit normals. 
If we were instead to put hard limits upon the minimum
value of $|x|$, choosing $p(x)$ to be uniform on $A<|x|<B$ where $A\neq 0$,  
then there would be a strict maximum on the spread $R \leq B/A$.
In our opinion, a hard minimum seems unnatural and we shall abandon the possibility.
Comparing the tails of ratio distributions obtained from normals and from the $[-3,3]$
uniform distributions then illustrates an important point: that a ratio
distribution's tail is not particularly sensitive to the detailed shape of
$p(x)$, provided that $p(x)$ remains non-negligible as $x \rightarrow 0$.

In truth, the probability of a large spread is more sensitive to the
\emph{measure}.
Had we abandoned the order-unity
assumption and instead taken the dimensionless couplings to be flat in
$\log_{10}|Y_i|$---the 
scale-invariant prior\footnote{Such a measure is mathematically equivalent to
the flat measure with a $1/|Y|$ prior.} which prefers no particular order of
magnitude --- the behaviour would change qualitatively: $\log_{10}R$ is then the
sample range of uniform variates, whose mean grows with the number of couplings
to fill whatever range is assumed, so that multi-decade hierarchies become
generic rather than rare. For two couplings, each varying over $a$ decades,
for 
instance, $P(R>1000)=(1-3/a)^2$, e.g.\ $0.49$ for $a=10$, against $0.001$ for
the unit normal flat-measure case. Such a scale-invariant prior would in fact describe the observed
quark and lepton masses reasonably well~\cite{Donoghue:2005cf}. We would
however wish
to explain the scale-invariance by the dynamics of some underlying model which
only has order-unity \emph{fundamental} dimensionless couplings $C_i$ from
which the scale-invariant prior is derived. We are therefore 
returned to the measure in the fundamental dimensionless couplings
rather than in their logarithm.
We shall also return to $p(x)$ being a unit normal as the default for the rest of
this paper, remembering from the argument above that the precise form of it
is not crucial as long as $p(x)$ is non-negligible for low values of $|x|$.  

There is one further step to obtain the probability distribution of the spread of
unit normal distributions:
the four preimages $\pm R$, $\pm 1/R$ of a given spread contribute equally,
since $t \rightarrow -t$ is a symmetry of (\ref{p_N}) and $1/t=Y_2/Y_1$ is
identically distributed by exchangeability of $Y_1$ and $Y_2$. 
Thus, we obtain the double-folded Cauchy distribution
\begin{equation}
  p_{\text{norm}}(R)=\frac{4}{\pi(1+R^2)},\qquad R\ge 1, \label{p_of_R_ans}
\end{equation}
whose tail still only falls\footnote{We note here that
a similar computation for the uniform prior gives $p_{\text{uni}}(R)=1/R^2$ at
large $R$.} like
$1/R^2$.
We can then compute
\begin{equation}
  P(R>r) = \int_r^\infty dR\ p_{\text{norm}}(R) = \frac{4}{\pi} \arctan (1/r).
  \end{equation}
Although the probability that $|Y_1|$ alone exceeds $10$ is a
minuscule $1.5\times10^{-23}$, the probability that the spread exceeds $10$ is
$P(R>10)=(4/\pi)\arctan(1/10)=0.127$: there is a substantial chance of an
order-of-magnitude ratio between two order-unity couplings.

We now extend the analysis from two order-unity couplings to $N$ of them. Since the spread
statistic depends only on the magnitudes of the couplings, we work directly with
$X_i := |Y_i|$. For a prior $p(x)$ that is symmetric about zero, the $X_i$ are
distributed on $[0,\infty)$ with PDF $g(x) =
2p(x)$ and cumulative distribution function (CDF)
\begin{equation}
  G(x) = \int_0^x g(x')\, dx' = \int_{-x}^{x} p(x')\, dx' = P({X_i} \leq x),
\end{equation}
where $P(C)$ denotes the probability of condition $C$.

It is instructive first to record the distribution of the largest coupling on
its own. Writing $M := \max_i X_i$ for the maximum and $F_M(x)$ for its CDF,
independence of the $X_i$ gives
\begin{eqnarray}
  F_M(x) &=& P(\max(X_1,\ldots,X_N) \leq x) \nonumber \\
  &=& P(X_1 \leq x, \ldots, X_N \leq x) \nonumber \\
  &=& \prod_{i=1}^N P(X_i \leq x) \;=\; G(x)^N,
\end{eqnarray}
with corresponding PDF $f_M(x) = dF_M/dx = N\, G(x)^{N-1} g(x)$.

The spread $R$, however, depends on both the largest coupling magnitude $M$ and the
smallest, $m := \min_i X_i$, through $R := M/m \geq 1$. We therefore need the
\emph{joint} distribution of $m$ and $M$. For $N$ IID variables, the joint PDF of the minimum
and the maximum is the standard order statistic
\begin{equation}
  f_{m,M}(a,b) = N(N-1)\, g(a)\, g(b)\, [G(b) - G(a)]^{N-2}, \qquad 0 \leq a \leq b,
  \label{eq:jointmM}
\end{equation}
where $a$ and $b$ are the realised values of $m$ and $M$ respectively: there are
$N$ ways to choose which coupling is the smallest and $N-1$ ways to choose the
largest, while each of the remaining $N-2$ couplings must lie in $[a,b]$,
contributing a factor $G(b) - G(a)$.

Changing variables from $(a,b)$ to $(a,R)$ with $b = Ra$ (so that the Jacobian is
$|\partial b/\partial R| = a$) and integrating over $a$ yields the spread PDF
\begin{equation}
  p_{\text{norm}}^{(N)}(R) = N(N-1) \int_0^\infty da\; a\, g(a)\, g(Ra)\, [G(Ra) - G(a)]^{N-2},
  \qquad R \geq 1.
  \label{eq:pRgen}
\end{equation}
For $N=2$ the bracket within the integrand is unity and (\ref{eq:pRgen}) reduces to
$p_\text{norm}^{(2)}(R) = 2\int_0^\infty a\, g(a) g(Ra)\, da$; with a unit normal $p(x)$ this
reproduces the double-folded Cauchy distribution (\ref{p_of_R_ans}).

For numerical work the CDF admits a more stable single-integral
form. Integrating (\ref{eq:jointmM}) over $b \geq ra$ and then over $a$ gives
\begin{equation}
  P(R>r) = 1 - N \int_0^\infty da\; g(a)\, [G(ra) - G(a)]^{N-1}.
  \label{eq:surv}
\end{equation}
We found no 
elementary closed form for (\ref{eq:surv}) and so we evaluate it
numerically and display various values in Table~\ref{tab:spread}. 
\begin{table}
  \begin{center}
  \begin{tabular}{|c|c|c|c|c|c|c|}
    \hline
    $N$ & 2&5 &10 &20 &50 &100 \\ \hline
    $r=10$ & 0.13 & 0.46 & 0.79 & 0.97 & 1.00 & 1.00 \\
    $r=100$ & 0.01 & 0.06 & 0.14 & 0.29 & 0.63 & 0.89 \\
    $r=1000$ & 0.001 & 0.006 & 0.015 & 0.034 & 0.095 & 0.196 \\
    \hline
  \end{tabular}
  \end{center}
  \caption{\label{tab:spread} The probability $P(R>r)$ that the spread of
    $N$ IID order-unity couplings exceeds $r$ for the unit normal prior, from
    (\ref{eq:surv}). The uncertainty in the numerical determination of the
    probabilities is smaller than the quoted precision.}
\end{table}
We see from the table that fairly large hierarchies are present with
non-negligible probabilities. If we take the
number of independent dimensionless couplings to be 20 (close to the number in
the SM) we see that, for example, there is a significant probability 
(because of the fat tails) that $R>100$: $P(R>100)=0.29$. 

This is one consequence of the fat tail at large $R$ seen above for
$N=2$ persisting at higher values of  
$N$. Substituting $a = s/R$ in 
(\ref{eq:pRgen}) and taking $R \gg 1$ gives 
\begin{equation}
p_{\text{norm}}^{(N)}(R)
\simeq C_N/R^2, \qquad  C_N = N(N-1)\, g(0) \int_0^\infty ds\; s\, g(s)\, G(s)^{N-2} ,
  \label{eq:CN}
\end{equation}
where $g(0) = \sqrt{2/\pi}$ for $p(x)$ being the unit normal.
Hence for fixed $N$ and large enough $r$, 
\begin{equation}
  P(R>r) \simeq C_N/r. \label{rR}
\end{equation}
Writing the PDF of the maximum of $N-1$ order-unity magnitude values,
\begin{equation}
  p_M^{(N-1)} =  (N-1) g(s) G(s)^{N-2},
\end{equation}
we can rewrite $C_N$ in the large $r$ limit as
\begin{equation}
C_N = N g(0) \langle M_{N-1} \rangle
  \Rightarrow
  P(R > r)  \simeq N g(0) \langle M_{N-1} \rangle / r, \label{prob}
\end{equation}
where
$\langle M_{N-1} \rangle = \int_0^\infty\ ds\; s\, p_M^{(N-1)}$ is the
expected value of the maximum of $N-1$ order-unity magnitude values. 
This only rises slowly with $N$;
rising from $0.80$ at 
$N=2$ to $2.5$ at
$N=50$. Ignoring this slow dependence, we see from (\ref{rR}) and 
(\ref{prob}) that\footnote{For large $N$, one can show that the probability of $R
> r$ tends to 1.}, for large $r$ and $C_N/r \ll 1$,
\begin{equation}
  P(R>r)\propto N / r.
\end{equation}

We now confront the order-unity hypothesis with the observed Standard Model 
charged-fermion masses. These arise from three independent $3 \times 3$ Yukawa
matrices $Y^u$, $Y^d$ and $Y^e$ with complex entries.
The Yukawa matrices couple the Higgs field to the up-type quarks,
down-type quarks and charged leptons; the physical masses are proportional to
their \emph{singular values} $\sigma_i$, since each Yukawa matrix
is diagonalised by a bi-unitary transformation\footnote{We set the neutrinos aside, as
their masses are most plausibly of seesaw origin and in any case lie beyond the
SM.}. Our order-unity hypothesis for such a matrix is
\emph{anarchy}~\cite{HallMurayamaWeiner:1999,HabaMurayama:2000}---structureless
entries, which we model as IID unit normals for both the real and the
imaginary parts of each entry. The spread\footnote{Here the spread
$R=\sigma_{\max}/\sigma_{\min}$ coincides with the $2$-norm
\emph{condition number} of the Yukawa matrix,
$\kappa_2(Y)=\|Y\|_2\,\|Y^{-1}\|_2=\sigma_{\max}/\sigma_{\min}$, where
$\sigma_{\max,\min}$ are the largest and smallest singular values; see
e.g.\ Ref.~\cite{Edelman:1988}; the identity
$\kappa_2(Y)=\sigma_{\text{max}}/\sigma_{\text{min}}$ is standard.} within a
sector 
(i.e.\ up quarks, down quarks or charged leptons) is then
\begin{equation}
  R=\frac{\sigma_{\max}}{\sigma_{\min}}=\frac{m_{\max}}{m_{\min}},
\end{equation}
which is
invariant under scaling the matrix by a factor: it is independent of the
overall coupling scale, 
so only the anarchy assumption enters.

\begin{figure}
  \begin{center}
    \includegraphics[width=\textwidth]{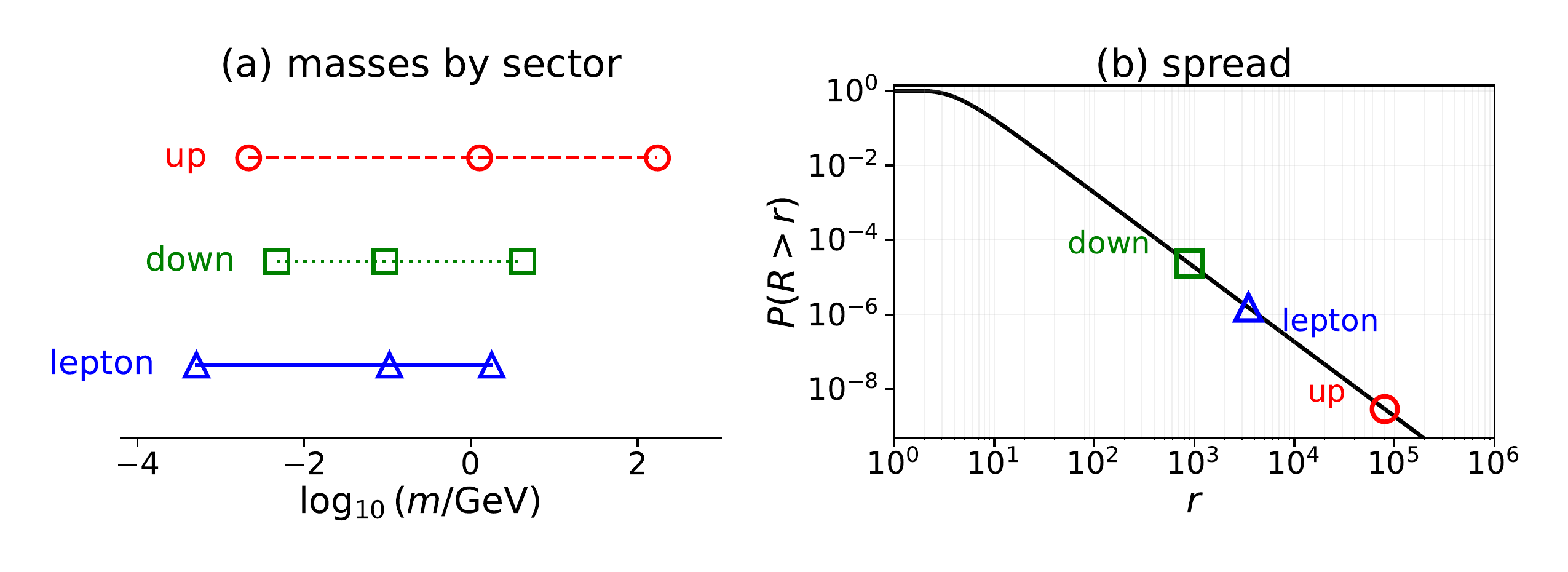}
  \end{center}
  \caption{\label{fig:sectors} (a) The SM charged-fermion masses, grouped by
  sector, on a logarithmic scale. (b) The survival function $P(R>r)$ of
  the spread of a $3\times3$ complex normal (anarchic) Yukawa matrix,
  (\ref{eq:condcdf}); the markers show the observed spread $R_{\rm obs}$ and
  the associated probability for each sector.}
\end{figure}
Crucially, the singular values are \emph{not} independent: they repel, and their
spread is governed by random-matrix theory~\cite{Wishart:1928,Forrester:2010,Edelman:1988} rather than by the order statistics of
IID couplings used above. As
derived in \ref{app:cond}, the spread  of a $3\times3$ 
matrix with normal complex entries has the closed-form CDF
\begin{equation}
  F_R(r)=\frac{2\,(r^2-1)^{8}\,\bigl(16r^4+31r^2+16\bigr)}
  {(r^2+2)^{5}\,(2r^2+1)^{5}}, \qquad (r\ge1),
  \label{eq:condcdf}
\end{equation}
a rational function of $r^2$ with $F_R(1)=0$ and $F_R(\infty)=1$. The
corresponding probability density function is
\begin{equation}
  p_R(r)=\frac{dF_R}{dr}
  =\frac{216\,r\,(r^2-1)^{7}\,(r^2+1)\,\bigl(11r^4+20r^2+11\bigr)}
  {(r^2+2)^{6}\,(2r^2+1)^{6}} .
  \label{eq:pdf}
\end{equation}
Expanding (\ref{eq:condcdf}) at large $r$ gives the tail
\begin{equation}
  P(R>r)=1-F_R(r)\;\xrightarrow[r\gg1]{}\;\frac{297}{16}\,\frac{1}{r^2}.
  \label{eq:tail}
\end{equation}
We display $P(R>r)$ in Fig.~\ref{fig:sectors}b. The median and mean of $R$
are respectively
\begin{align}
  R_{\rm med}&= 5.1,\\
  \langle R\rangle
  &=\frac{2131}{1024}
  +\frac{6825\sqrt2}{4096}\,\pi
  +\frac{4377\sqrt2}{2048}\,\arctan\!\sqrt2
  -\frac{5601\sqrt2}{1024}\,\arctan\!\frac{1}{\sqrt2}
  =7.6 .
  \label{eq:mean}
\end{align}
These values imply that anarchy typically spreads masses in a sector by
\emph{less than an 
order of magnitude}. 

The observed spreads in SM fermion masses are far
larger than the expectation given above, as shown in 
Fig.~\ref{fig:sectors}a. Table~\ref{tab:sectors} gives,  
for each sector, the ratio $R_{\rm obs}=m_{\max}/m_{\min}$ and the probability
$P(R>R_{\rm obs})$ from (\ref{eq:condcdf}). All three probabilities are
small; treating the matrices as independent, the
joint probability of reproducing all three spreads simultaneously is
tiny, being ${\mathcal O}(10^{-19})$. Correcting for renormalisation
effects does not change this outcome\footnote{Computing at the $Z^0$ boson mass
scale instead increases the probability by up to a factor of 2.}.
\begin{table}
  \begin{center}
  \begin{tabular}{|l|c|c|}
    \hline
    sector & $R_{\rm obs}$ & $P(R>R_{\rm obs})$ \\ \hline
    up-type quarks $(Y^u)$   & $8.0\times10^{4}$ & $2.9\times10^{-9}$  \\
    down-type quarks $(Y^d)$ & $8.9\times10^{2}$ & $2.3\times10^{-5}$  \\
    charged leptons $(Y^e)$  & $3.5\times10^{3}$ & $1.5\times10^{-6}$  \\ \hline
    combined                 &                   & $1.0\times10^{-19}$\\
    \hline
  \end{tabular}
  \end{center}
  \caption{\label{tab:sectors} For each charged-fermion sector, the observed
  largest-to-smallest mass ratio $R_{\rm obs}=m_{\max}/m_{\min}$ and the
  probability that an anarchic $3\times3$ unit normal complex Yukawa matrix has a spread at least this large, from the closed form (\ref{eq:condcdf}). We
  use modified minimal subtraction scheme
  ($\overline{\rm MS}$)
  light-quark masses at a
  renormalisation scale of $2$~GeV,
  heavy quarks (bottom and top)
  at their $\overline{\rm MS}$ masses and pole charged lepton
  masses~\cite{PDG}. The combined probability assumes that the three sectors
  are independent.} 
\end{table}
Anarchic order-unity Yukawa matrices are therefore very unlikely to account for
the observed charged-fermion hierarchies---sector by sector, and overwhelmingly
in combination. We note that in a successful grand unified theory, treating 
each sector as independent as we have done would be incorrect: some would in fact be
correlated (in $SU(5)$, for instance $Y^d=Y^e$ at leading order at a high
renormalisation scale), and so the combined probability would be much larger
(but still presumably at most $2.9\times 10^{-9}$, the probability of the up
sector). In any case, the results in Table~\ref{tab:sectors} confirm and
quantify the well-known defining feature of the charged fermion mass puzzle: that
the charged fermion masses show structure and are not likely drawn from an independent set
of order unity dimensionless couplings (anarchy).

\begin{figure}
  \begin{center}
    \includegraphics[width=0.5 \textwidth]{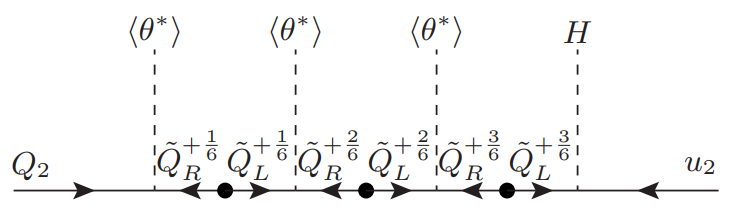}
  \end{center}
  \caption{\label{fig:spaghetti} An example \emph{spaghetti
  diagram}, taken from Ref.~\cite{Allanach:2019iiy}: a tree-level Feynman
    diagram by which an 
  effective charm-quark Yukawa coupling $(Y^u)_{22}$ is generated in an
  ultraviolet completion. 
  The external fermions $\psi^i_L,\psi^j_R$ and the SM Higgs doublet field $H$
  connect through a 
  chain of heavy vector-like quark fields $\tilde Q^X_{L,R}$ of masses $M_X$,
  taken to be of a 
  commone order ${\tilde M}$, i.e.\ $M_X \sim {\mathcal O}({\tilde M})$. Each 
  vertex ($\bullet$) carries a fundamental dimensionless coupling of order unity and each
  flavon insertion $\langle\theta\rangle$ a factor of order $\sim
  \epsilon:=\langle\theta\rangle/{\tilde M}$.
  Integrating out the heavy fields gives $|(Y^u)_{22}|=
  |\langle\theta\rangle^3 / (M_{1/6} M_{2/6} M_{3/6})|$ multiplied by four
  fundamental dimensionless couplings.} 
\end{figure}

This failure of anarchy at the level of the \emph{effective} Yukawa matrices
need not, however, indict the order-unity hypothesis for the truly fundamental
couplings.
In an ultraviolet (UV) completion such as the FN mechanism~\cite{Froggatt:1978nt}, the SM Yukawa couplings
are not all
themselves fundamental. In the FN mechanism, each entry $Y_{ij}$ is generated by tree-level chains of
heavy vector-like messenger exchanges---\emph{spaghetti
diagrams}~\cite{Allanach:1997sa}---every vertex of which carries a fundamental
coupling of order-unity, with the hierarchy set by the number of flavon (a
complex scalar SM singlet)
insertions through integer powers of a small
symmetry-breaking parameter $\epsilon$, see Fig.~\ref{fig:spaghetti}. The
number of fundamental 
couplings in the full UV theory, $N_{\rm UV}$, is typically large.
Two effects then widen the spread. First, as found above in (\ref{eq:CN}), the
spread of $N$ IID order-unity couplings grows with $N$, so a large $N_{\rm UV}$
already favours larger ratios. Second, and more importantly, an effective Yukawa
coupling is a \emph{product} of many such more fundamental couplings, and the
magnitude of a product of 
$\mathcal{O}(1)$ numbers spreads over a range that widens with the number of
factors---its logarithm being a sum that broadens like a random walk---so that
multiplicative combinations of anarchic couplings naturally generate hierarchical
spectra~\cite{vonGersdorff:2017}.
 The effective Yukawa couplings of
  lighter families tend to have more products of fundamental dimensionless
  couplings than the heavier families in FN models. Indeed, as illustrated
  numerically in Ref.~\cite{Greljo:2024zrj}, the effective Yukawa couplings of lighter families in FN models have a wider
  PDF than do the heavier families because of this.

  Thus, although a single anarchic $3\times3$
  matrix 
  only has a moderate spread, the effective SM Yukawa couplings assembled 
from many fundamental order-unity couplings can be far more hierarchical, in line
with observation in the SM. 
A quantitative analytic treatment of the resulting distribution of
spreads of effective Yukawa couplings, as a function of $N_{\rm UV}$ and the
FN charges, would also require a consideration of the independent
masses of vector-like representations of fermionic fields that appear in the
propagators of spaghetti
diagrams. We leave such considerations to future work. 

We note that Ref.~\cite{Hall:2007zj} analysed a large landscape of vacua in order to
describe fermion masses via the statistics of the landscape itself, where
extra dimensions provide a spectrum of couplings that result in a logarithmic
measure in supersymmetric theories. This reference contains
some (mostly numerical) statistical analysis of Yukawa couplings
within that context. Ref.~\cite{Babu:2016aro} used unit normal Yukawa couplings multiplied
by certain hierarchical factors in order to perform a different numerical study of the
statistics of fermion masses, including heavy fermionic vector-like representations of
the gauge group in models with unified symmetry and supersymmetry.
By contrast, we have provided a novel statistic to quantify adherence to the
order-unity expectation of a set of
generic dimensionless couplings: the spread $R$. We have obtained analytic results for
the distribution of $R$ for the first time in this context and
the spread of order-unity couplings is generically larger than the
naive expectation (that the spread itself be order-unity).
The matrix structure of the charged SM fermion Yukawa couplings quantitatively
changes the
spread of fermion masses and we have obtained 
an exact closed-form distribution of the spread in singular values of a
$3\times 3$ 
Yukawa matrix (with an IID complex unit normal for each entry),
agreeing with Refs.~\cite{24a9d2d23f354f5eb8ab41f308e696f8,PerezCastillo:2014}. 
The closed-form distribution (\ref{eq:condcdf}) quantifies the well-known fact that observed
hierarchies in the SM charged fermion masses are improbable from anarchy alone. 

\section*{Declaration of generative AI and AI-assisted technologies}
BCA has made use of {\tt Claude Opus 4.8} (Anthropic) to
produce some of the results reported here. All such results have been
independently checked by the author, who takes sole responsibility for the
content of this work.

\section*{Acknowledgements}
 This work has been partially supported by the Science and Technology Facilities Council consolidated grant
ST/X000664/1. We thank the members of the Cambridge Pheno Working Group for
helpful discussions, particularly P.~Chandra Sarjapur, who suggested to use
a variable similar to the spread in a different project about a specific FN
model. We also thank N.~Gubernari for helpful comments on the choice of prior
distribution. 


\appendix

\section{The spread of $3\times3$ normal matrices}

\label{app:cond}

In this appendix, we derive the distribution (\ref{eq:condcdf}).
For a $3\times3$ complex matrix $Y$
with IID unit complex normal\footnote{We remind the reader that a unit complex
normal $Y=(a +i b)/\sqrt{2}$
has both $a$ and $b$ drawn from IID real  unit normal distributions, so that
the variance of $|Y|$ is unity.} parts 
of each entry, the squared 
singular values $\ell_i=\sigma_i^2$ are the 
eigenvalues of the complex Wishart matrix $W=Y^{\dag}Y$~\cite{Wishart:1928};
for this square matrix case they form the Laguerre Unitary Ensemble~\cite{Forrester:2010}, with joint
density
\begin{equation}
  p(\ell_1,\ell_2,\ell_3)=\frac{1}{24}\,\big|\Delta(\ell)\big|^2
  \prod_{i=1}^3 e^{-\ell_i},\qquad 
  \Delta(\ell)=\prod_{i<j}(\ell_j-\ell_i),\ \ \ell_i\ge0 .
  \label{eq:lue}
\end{equation}
Note the Vandermonde factor $|\Delta(\ell)|^2$ encoding level repulsion.
The spread obeys $R\le r\iff\ell_{\max}\le r^2\ell_{\min}$. Writing
$\ell_i=\ell_{\min}u_i$ with $u_i\in[1,r^2]$ and integrating out the overall scale
$\ell_{\min}$ --- a Gamma integral, $\int_0^\infty\ell^{8}e^{-\ell
  s}\,d\ell=8!/s^{9}$ with $s=u_1+u_2+u_3$ where $u_1\equiv 1$ --- reduces the CDF to a two-dimensional integral,
\begin{equation}
  F_R(r)=10080\int_1^{r^2}\!\!\!\int_1^{u_3}
\frac{(u_2-1)^2(u_3-1)^2(u_3-u_2)^2}{(1+u_2+u_3)^{9}}
  \ du_2\,du_3,
  \label{eq:double}
\end{equation}
the prefactor being fixed by $F_R(\infty)=1$.
The inner integral over $u_2$ and the outer one over $u_3$ are elementary
rational integrals,
yielding the closed form (\ref{eq:condcdf}).

An analogous computation to the one presented above leads to the following
closed-form CDF of the spread of a $3 \times 3$ matrix with \emph{real} unit normal entries,
which we include for completeness
\begin{equation}
  F_R(r)=%
\frac{(r^2-1)^2}{(r^2+1)^3} \left[ \frac{r(r^4 + 8 r^2 + 11)}{(r^2+2)^{3/2}}-
      \frac{11r^4 + 8r^2 + 1}{(2 r^2 + 1)^{3/2}}\right],
  \label{eq:condcdfreal}
\end{equation}
which implies
\begin{equation}
 P(R>r) \begin{array}{c}\longrightarrow \\ r\gg 1\\ \end{array} \frac{11\sqrt2}{4} \frac{1}{r}.
 \label{eq:condtail}
\end{equation}
The real coupling matrix is thus more poorly conditioned with a fatter tail than the
complex case (where $P(R>r) \propto 1/r^2$ at large $r$).
The real case has no well-defined mean of $R$ (in contrast to the complex case),
although its median $R=7.1$ is comparable to the complex case's median
$R=5.1$. The real $3 \times 3$ matrix case could be relevant for specific beyond the SM theories
where a symmetry enforces the entries of a $3 \times 3$ mixing matrix to be real. 

  \bibliographystyle{JHEP} 
  \bibliography{order_one}

@article{Greljo:2024zrj,
    author = "Greljo, Admir and Thomsen, Anders Eller and Tiblom, Hector",
    title = "{Flavor hierarchies from SU(2) flavor and quark-lepton unification}",
    eprint = "2406.02687",
    archivePrefix = "arXiv",
    primaryClass = "hep-ph",
    doi = "10.1007/JHEP08(2024)143",
    journal = "JHEP",
    volume = "08",
    pages = "143",
    year = "2024"
}

@article{Allanach:2019iiy,
    author = "Allanach, B. C. and Davighi, Joe",
    title = "{Naturalising the third family hypercharge model for neutral current $B$-anomalies}",
    eprint = "1905.10327",
    archivePrefix = "arXiv",
    primaryClass = "hep-ph",
    doi = "10.1140/epjc/s10052-019-7414-z",
    journal = "Eur. Phys. J. C",
    volume = "79",
    number = "11",
    pages = "908",
    year = "2019"
}

@article{Hall:2007zj,
    author = "Hall, Lawrence J. and Salem, Michael P. and Watari, Taizan",
    title = "{Statistical Understanding of Quark and Lepton Masses in Gaussian Landscapes}",
    eprint = "0707.3446",
    archivePrefix = "arXiv",
    primaryClass = "hep-ph",
    reportNumber = "UCB-PTH-07-12, LBNL-62798, CALT-68-2654, UT-07-19",
    doi = "10.1103/PhysRevD.76.093001",
    journal = "Phys. Rev. D",
    volume = "76",
    pages = "093001",
    year = "2007"
}

@article{Babu:2016aro,
    author = "Babu, K. S. and Khanov, Alexander and Saad, Shaikh",
    title = "{Anarchy with Hierarchy: A Probabilistic Appraisal}",
    eprint = "1612.07787",
    archivePrefix = "arXiv",
    primaryClass = "hep-ph",
    reportNumber = "OSU-HEP-16-10",
    doi = "10.1103/PhysRevD.95.055014",
    journal = "Phys. Rev. D",
    volume = "95",
    number = "5",
    pages = "055014",
    year = "2017"
}

@article{Witten:1985xb,
    author = "Witten, Edward",
    title = "{Dimensional Reduction of Superstring Models}",
    reportNumber = "Print-85-0244 (PRINCETON)",
    doi = "10.1016/0370-2693(85)90976-1",
    journal = "Phys. Lett. B",
    volume = "155",
    pages = "151",
    year = "1985"
}

@article{Kaplunovsky:1995jw,
    author = "Kaplunovsky, Vadim and Louis, Jan",
    title = "{On Gauge couplings in string theory}",
    eprint = "hep-th/9502077",
    archivePrefix = "arXiv",
    reportNumber = "UTTG-24-94, LMU-TPW-94-24",
    doi = "10.1016/0550-3213(95)00172-O",
    journal = "Nucl. Phys. B",
    volume = "444",
    pages = "191--244",
    year = "1995"
}

@article{TOE,
    author = "Ellis, John",
    title = "The superstring: theory of everything, or of nothing?",
    doi = "10.1038/323595a0",
    journal = "Nature",
    volume = "323",
    pages = "595-598",
    year = "1986"
}

@article{PDG,
    author = "Navas, S. and others",
    collaboration = "Particle Data Group",
    title = "{Review of Particle Physics}",
    journal = "Phys. Rev. D",
    volume = "110",
    number = "3",
    pages = "030001",
    year = "2024",
    doi = "10.1103/PhysRevD.110.030001"
}

@article{Froggatt:1978nt,
    author = "Froggatt, C. D. and Nielsen, H. B.",
    title = "{Hierarchy of Quark Masses, Cabibbo Angles and CP Violation}",
    journal = "Nucl. Phys. B",
    volume = "147",
    pages = "277--298",
    year = "1979",
    doi = "10.1016/0550-3213(79)90316-X"
}

@article{Donoghue:2005cf,
    author = "Donoghue, John F. and Dutta, Koushik and Ross, Andreas",
    title = "{Quark and lepton masses and mixing in the landscape}",
    eprint = "hep-ph/0511219",
    archivePrefix = "arXiv",
    doi = "10.1103/PhysRevD.73.113002",
    journal = "Phys. Rev. D",
    volume = "73",
    pages = "113002",
    year = "2006"
}

@article{Geary:1930,
    author = "Geary, R. C.",
    title = "{The Frequency Distribution of the Quotient of Two Normal Variates}",
    journal = "J. Roy. Statist. Soc.",
    volume = "93",
    number = "3",
    pages = "442--446",
    year = "1930",
    doi = "10.2307/2342070"
}

@article{HallMurayamaWeiner:1999,
    author = "Hall, Lawrence J. and Murayama, Hitoshi and Weiner, Neal",
    title = "{Neutrino mass anarchy}",
    eprint = "hep-ph/9911341", archivePrefix = "arXiv",
    journal = "Phys. Rev. Lett.", volume = "84", pages = "2572--2575", year = "2000",
    doi = "10.1103/PhysRevLett.84.2572"
}

@article{HabaMurayama:2000,
    author = "Haba, Naoyuki and Murayama, Hitoshi",
    title = "{Anarchy and hierarchy}",
    eprint = "hep-ph/0009174", archivePrefix = "arXiv",
    journal = "Phys. Rev. D", volume = "63", pages = "053010", year = "2001",
    doi = "10.1103/PhysRevD.63.053010"
}

@article{vonGersdorff:2017,
    author = "von Gersdorff, Gero",
    title = "{Natural Fermion Hierarchies from Random Yukawa Couplings}",
    eprint = "1705.05430", archivePrefix = "arXiv",
    journal = "JHEP", volume = "09", pages = "094", year = "2017",
    doi = "10.1007/JHEP09(2017)094"
}

@article{Edelman:1988,
    author = "Edelman, Alan",
    title = "{Eigenvalues and condition numbers of random matrices}",
    journal = "SIAM J. Matrix Anal. Appl.", volume = "9", number = "4",
    pages = "543--560", year = "1988", doi = "10.1137/0609045"
}

@article{PerezCastillo:2014,
    author = "P\'erez Castillo, Isaac and Katzav, Eytan and Vivo, Pierpaolo",
    title = "{Phase transitions in the condition-number distribution of Gaussian random matrices}",
    eprint = "1403.1185", archivePrefix = "arXiv",
    journal = "Phys. Rev. E", volume = "90", number = "5", pages = "050103", year = "2014",
    doi = "10.1103/PhysRevE.90.050103"
}

@article{Wishart:1928,
    author = "Wishart, John",
    title = "{The generalised product moment distribution in samples from a normal multivariate population}",
    journal = "Biometrika", volume = "20A", pages = "32--52", year = "1928",
    doi = "10.1093/biomet/20A.1-2.32"
}

@book{Forrester:2010,
    author = "Forrester, Peter J.",
    title = "{Log-Gases and Random Matrices}",
    publisher = "Princeton University Press", year = "2010"
}

@inproceedings{24a9d2d23f354f5eb8ab41f308e696f8,
title = "Exact expressions for the condition number distribution of complex Wishart matrices",
author = "Michail Matthaiou and M.R. McKay and Smith, \{Peter J\} and J.A. Nossek",
year = "2010",
language = "English",
booktitle = "Proc. IEEE International Conference on Communications (ICC)",
publisher = "Institute of Electrical and Electronics Engineers Inc.",
address = "United States",
}

@article{Allanach:1997sa,
    author = "Allanach, B. C. and King, S. F.",
    title = "{String unification, spaghetti diagrams and infrared fixed points}",
    eprint = "hep-ph/9703293", archivePrefix = "arXiv",
    journal = "Nucl. Phys. B", volume = "507", pages = "91", year = "1997"
}





\end{document}